\documentclass[]{interact}
\usepackage{epstopdf}
\usepackage[caption=false]{subfig}
\usepackage[numbers,sort&compress]{natbib}
\usepackage{hyperref}
\usepackage{multirow}
\usepackage{lmodern}
\usepackage{booktabs}
\usepackage{url}
\usepackage{amsmath,amssymb,amsfonts}
\usepackage{graphicx}
\usepackage{textcomp}
\usepackage{tikz}
\usepackage{algorithm}
\usepackage{algorithmic}
\usepackage{array}

\bibpunct[, ]{[}{]}{,}{n}{,}{,}
\renewcommand\bibfont{\fontsize{10}{12}\selectfont}
\makeatletter
\def\NAT@def@citea{\def\@citea{\NAT@separator}}
\makeatother

\theoremstyle{plain}

\theoremstyle{definition}

\theoremstyle{remark}

\begin{document}

\articletype{ARTICLE TEMPLATE}

\title{VisiTrail: A Cognitive Visualization Tool for Time-Series Analysis of Eye Tracking Data from Attention Game}

\author{
\name{Abdul Rehman\textsuperscript{a}, Ilona Heldal\textsuperscript{a}, and Jerry Chun-Wei Lin\textsuperscript{a}\thanks{CONTACT: Abdul Rehman. Email: \href{mailto:arj@hvl.no}{arj@hvl.no}}}
\affil{\textsuperscript{a}Department of Computer Science, Electrical Engineering and Mathematical Sciences,\\
Western Norway University of Applied Sciences, Bergen, Norway}
}

\maketitle

\begin{abstract}
Eye Tracking (ET) can help to understand visual attention and cognitive processes in interactive environments. In attention tasks, distinguishing between relevant target objects and distractors is crucial for effective performance, yet the underlying gaze patterns that drive successful task completion remain incompletely understood. Traditional gaze analyses lack comprehensive insights into the temporal dynamics of attention allocation and the relationship between gaze behavior and task performance. When applied to complex visual search scenarios, current gaze analysis methods face several limitations, including the isolation of measurements, visual stability, search efficiency, and the decision-making processes involved in these scenarios. This paper proposes an analysis tool that considers time series for eye tracking data from task performance and also gaze measures (fixations, saccades and smooth pursuit); temporal pattern analysis that reveals how attention evolves throughout task performance; object-click sequence tracking that directly links visual attention to user actions; and performance metrics that quantify both accuracy and efficiency. This tool provides comprehensive visualization techniques that make complex patterns of stimuli and gaze connections interpretable. 
\end{abstract}


\section{Introduction}
Eye tracking is a widely utilized technology for analyzing how individuals visually interact with digital environments through gaze measures \cite{rehman2024towards,novak2024eye,ali2023towards}. By calculating where and for how long someone focuses on different areas of a screen, one can gain valuable insights into attention, decision-making processes, and cognitive load \cite{daehlen2024towards}. Recently, using ET technology has seen significant application in serious game-based scenarios, where users engage dynamically with evolving visual representations \cite{costescu2023mushroom}. Unlike static images, games invite variation in visual attention during gameplay \cite{costescu2023mushroom}, which can be recorded and measured by gaze data collected by ET. Analyzing gaze changes over time, along with the progression of games and interactions, is referred to as temporal analysis \cite{lamsa2022focus}. It can enable us to determine not just the areas where people direct their gaze but also approximate attention transitions throughout the gaming experience \cite{keshava2024just}.

Examining gaze behavior changes within a temporal framework along with fixation patterns, saccadic movements, and response times from interactive tasks in serious games offers deep insights that can enable (educators and psychologists do not manage this yet) to quantify attention deficits, evaluate the effectiveness of the intervention, and personalize learning experiences based on individual cognitive profiles \cite{papavlasopoulou2021investigating}. For instance, identifying how quickly a player notices critical game elements or how their focus adapts to new challenges can inform both game development and the evaluation of gameplay. Conventional gaze analysis often emphasizes aggregated data such as heatmaps (intensity and frequency of fixation), which may overlook vital time-sensitive dynamics. Adopting a chronological perspective enables researchers to uncover patterns, such as hesitation, learning curves, or shifts in focal areas, which static summaries often fail to capture. This form of analysis is particularly beneficial in serious games or training simulations, where tracking attention flow could help to evaluate performance or engagement levels with greater precision and depth. To overcome the challenges above, this paper makes the following contributions:
 
\begin{itemize}
\item We propose a comprehensive cognitive visualization tool named VisiTrail for time-series analysis tool that integrates multiple analytical dimensions: enhanced movement classification that distinguishes between fixations, saccades, and smooth pursuit; temporal pattern analysis that reveals how attention evolves throughout task performance; object-click sequence tracking that directly links visual attention to user actions; and performance metrics that quantify both accuracy and efficiency of visual search behavior. 
\item We provide a comprehensive tool that makes complex patterns interpretable to researchers and practitioners, allowing them to create personalized strategies that align with the user's natural attention rhythms.
\item The source code and demonstration video of VisiTrail are available \footnote{https://github.com/arnor-git/VisiTrail}
\end{itemize}

The rest of the paper is organized as follows. Section \ref{Related} discusses related work on eye tracking and learning environments. Section \ref{Proposed} presents our proposed approach for enhancing the Understanding of Time-Series Analysis of Eye-Tracking Data in interactive learning systems. Section \ref{results} describes the experimental setup and results, including privacy and performance evaluation. Finally, Section \ref{conclusion} concludes the paper and outlines possible directions for future research.

\section{Related Work}\label{Related}
Frutos-Pascual et al. \cite{frutos2015assessing} utilized ET technology to identify children's behavior in attention-enhancement therapies. By analyzing eye movement patterns during interaction with puzzle games, the researchers found that participants with better performance exhibited quantifiably different eye movement patterns compared to those with poorer results. Piazzalunga et al. \cite{piazzalunga2023investigating} used serious games-based ET data to identify children at risk of dysgraphia. By analyzing scan paths and fixation patterns during gameplay, the study found that children with poorer performance had chaotic scan paths.
In contrast, those who performed better had more ordered scan paths. The integration of ET metrics with game performance data provided a nuanced understanding of visual perception impairments. Velichkovsky et al. \cite{velichkovsky2019visual} analyzed the distributions of fixation durations among professional, amateur, and novice eSports players. They found that highly skilled gamers exhibit more variability in fixation durations and bimodal distribution patterns, indicating the presence of both ambient and focal fixation types. 

Kim et al.~\cite{kim2024development} aimed to develop a deep learning (DL) model to identify individuals with mental illnesses who have impaired visuospatial memory encoding. During a 3-minute memorization test of the RCFT, eye movements were recorded to evaluate the structure and retention of psychosis, obsessive-compulsive disorder (OCD), and responses from healthy controls. The resulting scores and fixation points, which show areas of eye focus, were used to create a Long Short-Term Memory (LSTM) model with an attention mechanism designed to distinguish between normal and impaired executive function. Argasinski et al. \cite{argasinski2017patterns} provided a framework for developing and evaluating serious games that incorporate ET and biosensor data, suggesting that the inclusion of temporal data is better for assessing user interactions and affective responses during gameplay. Hajari et al. \cite{hajari2018spatio} explored team cognition by analyzing spatio-temporal ET data during laparoscopic simulation operations. Using Cross Recurrence Analysis (CRA) and overlap analysis, they identified features that distinguish between high- and low-performing teams based on temporal gaze data patterns, thereby enhancing understanding of collaborative performance. 

Du et al.~\cite{du2024privategaze} introduced PrivateGaze, a solution that protects users' private information while interacting with black-box gaze tracking data, all while maintaining estimation accuracy. PrivateGaze effectively preserves sensitive user data, such as identity and gender, as demonstrated through experiments conducted on four benchmark datasets. Paskovske et al. \cite{paskovske2024eye} utilized ET to analyze how participants distribute their attention. Their findings revealed a notable difference between novices and experts based on the number of fixations. Experts tend to spend less time on tasks and employ more efficient problem-solving strategies. Additionally, the prevalence of NDDs has been increasing globally, leading to a greater number of children with disabilities being integrated into mainstream schools. This shift calls for equitable and appropriate treatment of learners with disabilities, driven by the implementation of educational policies and practices that support inclusive education.

In the past, Costescu et al. \cite{costescu2023mushroom} primarily focused on developing a platform to enhance children's play experiences. A key concept of this platform is identifying the focus of children's attention during play activities. This can be examined through the game "Mushroom Hunter," which assesses the ability to sustain attention. Bueno et al. \cite{bueno2023datasets} focused on developing an application specifically designed for AI research in education, particularly for children with NDD. Thill et al., \cite{thill2022modelling} investigated how robotics and social AI interpret children's behavior. Additionally, research by other studies \cite{daehlen2024towards,rehman2024towards} explored how serious games, combined with eye-tracking technology, can provide insights that help teachers better support children with NDD. The study \cite{costescu2020development} proposed here builds on research into the development of platforms designed to assist children with NDD. This study focuses on providing a time-series analysis tool that integrates multiple analytical dimensions. It includes enhanced movement classification, which distinguishes between fixations, saccades, and smooth pursuit. The tool also offers temporal pattern analysis to reveal how attention evolves during task performance. Additionally, it features object-click sequence tracking that directly links visual attention to user actions. Performance metrics quantify both the accuracy and efficiency of visual search behavior, making complex patterns interpretable to researchers and practitioners. This allows them to develop personalized strategies that align with the user’s natural attention rhythms.

\section{VisiTrail (Proposed Tool)}\label{Proposed}
\begin{figure*}[!ht]
 \centering
 \includegraphics[width=0.9\textwidth]{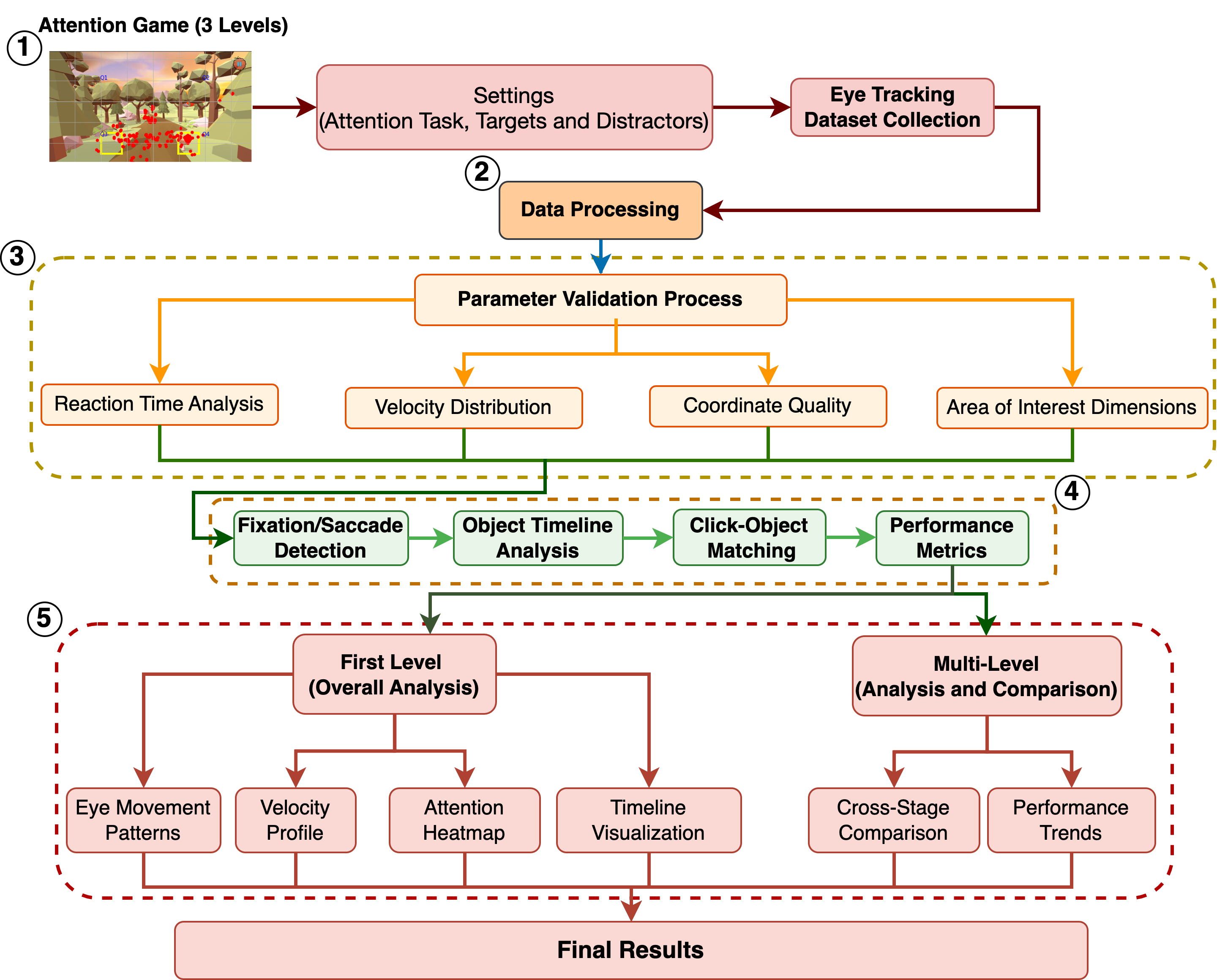}
 \caption{Proposed VisiTrail tool for Time-Series Analysis of Attention Game based Eye-Tracking Data}
 \label{figpropsod}
\end{figure*}

Fig. \ref{figpropsod} provides an overview of the proposed tool from data collection, processing, quality validation, parameter selection, gaze classification, and performance metrics for two scenarios: a first-level overall analysis to provide detailed insights and a multilevel analysis to compare performance across all three levels.
 
Algorithm~\ref{alg:eyetrack} implements a comprehensive eye-tracking analysis pipeline designed to enhance the understanding of time-series analysis of eye-tracking data. The process begins by parsing coordinate strings from CSV format, followed by the removal of incomplete records and validation of gaze positions within screen boundaries, using a tolerance of 50 pixels. Next, gaze behavior is classified using velocity-based thresholds. We select a threshold of 721 px/s that differentiates fixation, indicating focused attention, from saccades, which are rapid eye movements occurring during visual searches calculated by using the I-VT algorithm \cite{salvucci2000ivt,olsen2012tobii}. Following this, the algorithm employs greedy target-click matching with temporal constraints. A minimum reaction time of 522 ms helps prevent false matches from anticipatory responses, while a maximum time of 5000 ms ensures that response windows remain realistic. The algorithm quantifies task performance through several metrics: the hit rate (indicating target detection accuracy), the false alarm rate (reflecting attention control), and the average reaction time (measuring processing speed). It also computes screen utilization and gaze path length to assess visual search strategies and patterns of attention distribution. Finally, the algorithm generates evidence-based recommendations for educators and psychologists based on validated performance thresholds derived from developmental and clinical guidelines.

\begin{algorithm}[!ht]
\caption{Algorithm for Time-Series Analysis of Attention Game based Eye-Tracking Data (VisiTrail Tool)}
\label{alg:eyetrack}

\begin{algorithmic}[1]
\STATE \textbf{Input:} Mushroom Game (Attention) Dataset
\STATE \textbf{Parameters:} $V_{thresh} = 721$ px/s, $RT_{min} = 522$ ms, $RT_{max} = 5000$ ms
\STATE \textbf{Pre-processing:} Parse coordinates, remove missing data, filter valid screen bounds
\STATE \textbf{I-VT Fixation Detection:}
\FOR{$i = 2$ to $n$}
    \STATE $v_{mag,i} \leftarrow \frac{\sqrt{(x_i - x_{i-1})^2 + (y_i - y_{i-1})^2}}{(Timestamp_i - Timestamp_{i-1})/1000}$
    \STATE $movement\_type_i \leftarrow$ fixation if $v_{mag,i} \leq V_{thresh}$, else saccade
\ENDFOR
\STATE \textbf{Event Extraction:} Extract appear/disappear events and clicks, sort by timestamp
\STATE \textbf{Response Matching:}
\STATE $T_{targets} \leftarrow$ target appear events, $C_{correct} \leftarrow$ correct clicks
\STATE Initialize $matched\_pairs \leftarrow 0$, $used\_clicks \leftarrow \emptyset$
\FOR{each target $t \in T_{targets}$}
    \STATE Find earliest unused click $c$ where $t.time < c.time$ and $c \notin used\_clicks$
    \STATE $rt \leftarrow c.timestamp - t.timestamp$
    \IF{$RT_{min} \leq rt \leq RT_{max}$}
        \STATE $matched\_pairs \leftarrow matched\_pairs + 1$, $used\_clicks \leftarrow used\_clicks \cup \{c\}$
    \ENDIF
\ENDFOR
\STATE \textbf{Metrics:} Compute hit rate, false alarm rate, reaction times, screen utilization
\STATE \textbf{Recommendations:} Generate Recommendations based on performance thresholds
\STATE \textbf{Output:} Performance metrics, spatial metrics, recommendations
\end{algorithmic}
\end{algorithm}

\subsection{Dataset Selection, Pre-Processing and Data Quality Validation}
Costescu et al.~\cite{costescu2023mushroom} created and structured the sustained attention game that was used to gather data. It implements the Computerized Continuous Performance Task (CCPT) according to the original task design created by~\cite{CCPT}. Participants were shown a forest road with flowers (\textit{distractors}) and mushrooms (\textit{targets}) sporadically appearing on either side. The goal is to avoid interacting with distractions and touch the screen when a target appears. Data was gathered using a Tobii Pro Nano eye-tracker \footnote{\url{https://connect.tobii.com/s/article/how-to-configure-your-tobii-pro-nano-x2-or-x3-eye-tracker-with-the-mobile-device-stand?language=en_US}}, and the game utilizes a head-positioning and calibration screen specifically designed for the target demographic. We specifically use Pilot 3 data collected from Romania, which includes ET data from 8 students at a special education school in Romania~\cite{costescu2023mushroom}. The first step in the eye-tracking scoring system is to clean and prepare raw input data for difficulty levels 1, 2, and 3. The system begins by loading CSV files to extract eye position coordinates, which are initially stored as text strings (e.g., "(1250, 680)"). These strings are converted into numerical values for analysis, and object positions and area-of-interest boundaries are extracted similarly. Next, the data is cleaned by removing invalid entries, such as records where the eye tracker detected zero coordinates or where tracking was inaccurate. Incomplete records that could cause errors in analysis are also filtered out. The preprocessing phase adds necessary metadata, including student labels, difficulty levels, and normalized timestamps, to ensure proper sequencing and data integrity. Finally, the three-level files are combined into one comprehensive dataset. After preprocessing, the cleaned dataset contains valid eye-tracking coordinates, game interaction events, and correctly labeled metadata, all of which are ready for detailed analysis.

In the data quality validation step, we verified that the ET data was accurate and removed any incorrect information. Eye trackers sometimes record impossible locations, such as negative coordinates or positions far outside the computer screen, so we needed to clean this up before analyzing the data. We further removed the rows that had missing ET data. There were 304 Initial instances, and 30 of them were removed.

\subsection{Parameter Validation Process}
The parameter validation level represents the methodological basis of this ET analysis tool, employing a sophisticated, data-driven approach that fundamentally departs from conventional assumption-based methodologies typically used in cognitive assessment research. We focus on four levels, especially reaction time analysis, velocity distribution, coordinate quality and the area of interaction dimension. For reaction time, we measured the time it took for students to respond to an object's appearance by recording the moment the object appeared and when the student clicked on it. We found that most responses occurred between 522 milliseconds and 5000 milliseconds, so we used these as our minimum and maximum reaction times. Anything faster was likely an accident, and anything slower indicated that the student was not paying attention. For velocity distribution analysis, we measured the distance between the student's eyes at one moment and the next, then divided by the time difference to calculate the speed \cite{salvucci2000ivt,holmqvist2011eye}. We tested different speed limits and found that 721 pixels per second was the best cutoff. Any value lower meant the eyes were staying still (fixation), while anything faster meant the eyes were jumping around (saccade) \cite{henderson2003human}. Our threshold selection was not arbitrary but based on systematic empirical validation of our eye-tracking data.

\begin{figure}[!ht]
\centering
\includegraphics[trim={0 0 0 0.9cm},clip,width=1\linewidth]{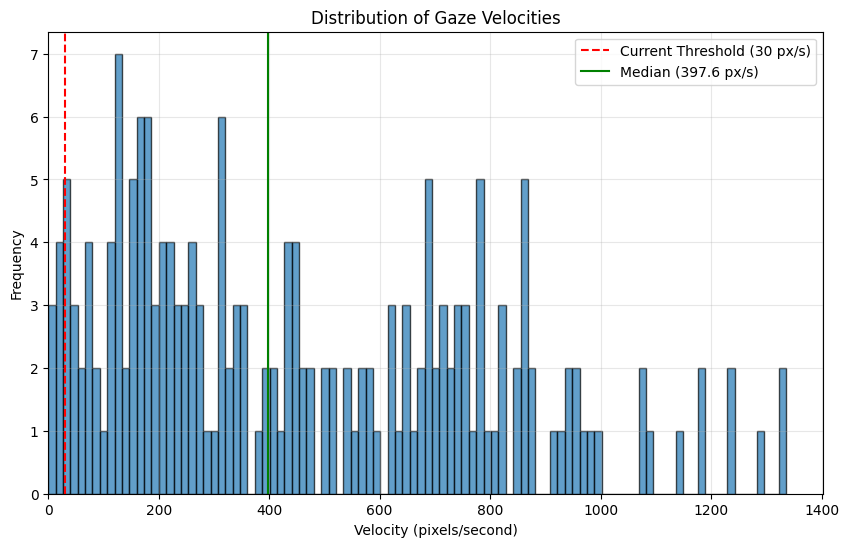}
\caption{Distribution of gaze velocities (in pixels/second)}
\label{fig:Distribution}
\end{figure}

The threshold for classifying fixations was set at 721 pixels per second, based on the 75th percentile of the gaze velocity distribution after removing outliers, as illustrated in Fig. \ref{fig:Distribution}. This threshold indicates that 75\% of the recorded gaze velocities fall below this value, allowing us to conservatively identify slower movements as fixations, while filtering out faster movements, which are likely saccades. The analysis revealed that commonly used lower thresholds, such as 30 pixels per second, captured only a small fraction of the data (approximately 4.2\%) as fixations, significantly underestimating the actual number of fixations in this attention game dataset. By using the 75th percentile value, we align the threshold with the natural distribution of the data, enhancing the realism and accuracy of fixation detection in the context of IVT analysis. The resulting velocity histogram and fixation rates support this choice, demonstrating that 721 pixels per second effectively distinguishes fixations from rapid eye movements in this specific dataset.

 For coordinate quality analysis, we tested the strictness of ET accuracy by examining the number of data points that fell outside the screen boundaries. We then tried different tolerance levels to determine which one retained the most accurate data while removing obvious mistakes \cite{nyström2010detection}. The 50-pixel coordinate tolerance strikes a balance between data retention and tracking accuracy, preserving eye-tracking points while excluding clear measurement errors. Target success rates use temporal matching only, pairing targets chronologically with subsequent correct clicks within the reaction time window without spatial distance validation. 
\subsection{Gaze Classification Analysis}
The level used all validated parameters from the previous level to understand the student's behavior throughout the gameplay. This level had four main steps: fixation/saccade detection \cite{henderson2003human}, object timeline analysis \cite{jarodzka2010eye}, click-object matching, and performance metrics. First, we classified the students' eye movements by examining the speed of their eye movements at each moment. Next, we created a timeline of all the events that occurred during gameplay by reviewing the data and marking the instances when objects appeared on the screen, disappeared, and when the student clicked on them. We then arranged these events in chronological order to visualize the sequence of events. After that, we matched up clicks with objects by looking at each time an object appeared and checking if the student clicked somewhere nearby within a reasonable amount of time; we used our validated settings (522-5000 milliseconds for timing) to decide if a click was meant for a specific object or if it was just random. Finally, we calculated performance metrics by counting the number of targets the student successfully clicked, the number of mistakes they made on distractors, the speed of their reactions, and the accuracy of their clicks, providing us with numbers that indicated how well the student was paying attention and responding during gameplay. 
 
After analyzing the data, we examined two scenarios: the first scenario presented the overall analysis, providing a detailed breakdown of each student, and the multilevel analysis compared the performance of each student across all three levels. We create four different types of visual displays: (1) a timeline chart showing exactly when objects appeared and when clicks happened, (2) an eye movement pattern map showing where the student looked and how their eyes moved around the screen, (3) a velocity graph showing when their eyes were moving fast or slow over time, an attention heatmap with colored areas showing where they spent the most time looking plus markers for their clicks, and (4) a performance dashboard with charts showing their success rates and reaction times. For multilevel analysis, we created comparison charts that show how students' performance changed from one level to the following, trend graphs that indicate whether they improved or worsened over time, and summary tables that display all the essential numbers side by side for easy comparison.

\section{VisiTrail Analysis and Results}\label{results}
\begin{figure*}[!ht]
    \centering
    \begin{minipage}[t]{0.8\textwidth}
        \centering
        \includegraphics[width=\linewidth]{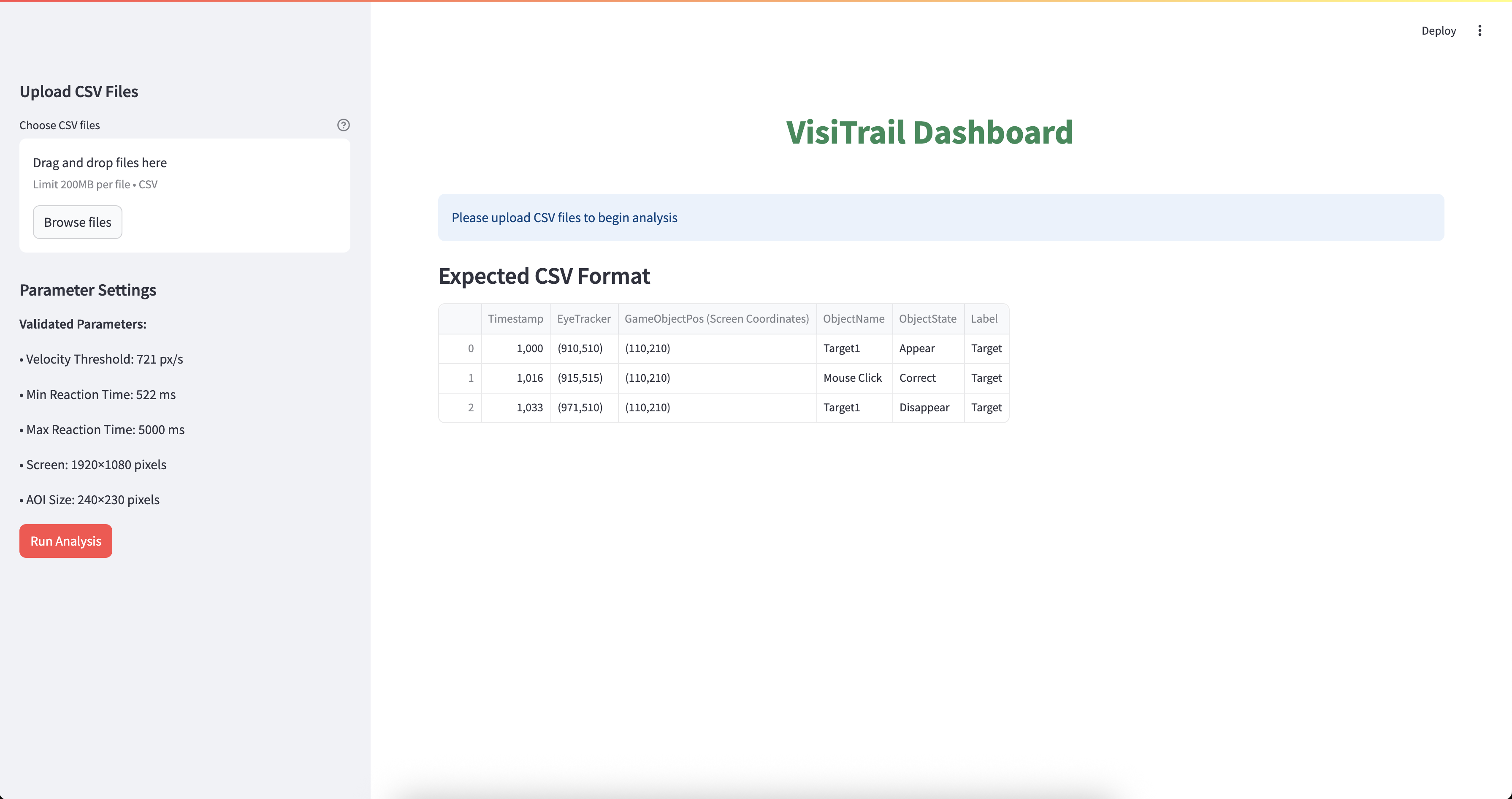}
        \caption{GUI of the VisiTrail (Sample 1)}
        \label{GUI1}
    \end{minipage}
    \begin{minipage}[t]{0.8\textwidth}
        \centering
        \includegraphics[width=\linewidth]{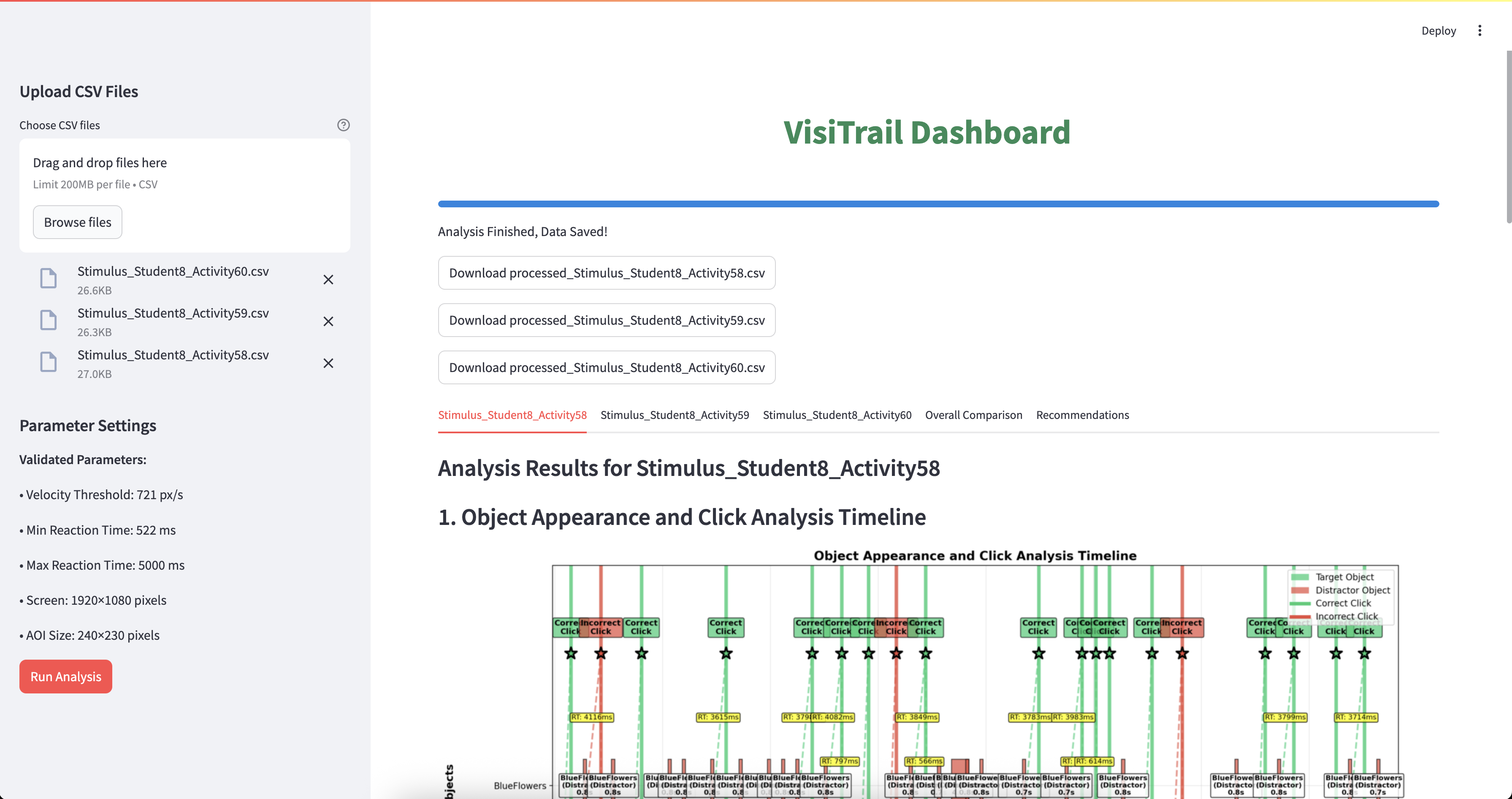}
        \caption{GUI of the VisiTrail (Sample 2)}
        \label{GUI2}
    \end{minipage}
\end{figure*}

This section demonstrates the results of the proposed tool. Figs. \ref{GUI1}, \ref{GUI2} depict the Graphical User Interface (GUI) of the proposed VisiTrail Tool. The GUI initially displays the expected data format and predefined parameters. Further, it asks the user to input the attention game data. After inputting the data, it generates 5 tables: the first three for individual game levels, the 4th one for the overall comparison, and the 5th one for the recommendation. For this study, we focus on only one student's data (Student 8), as the purpose was to gain insight into the student's behavior.

\begin{figure*}[!ht]
 \centering
 \includegraphics[trim={0 0 0 0.95cm},clip,width=1.03\textwidth]{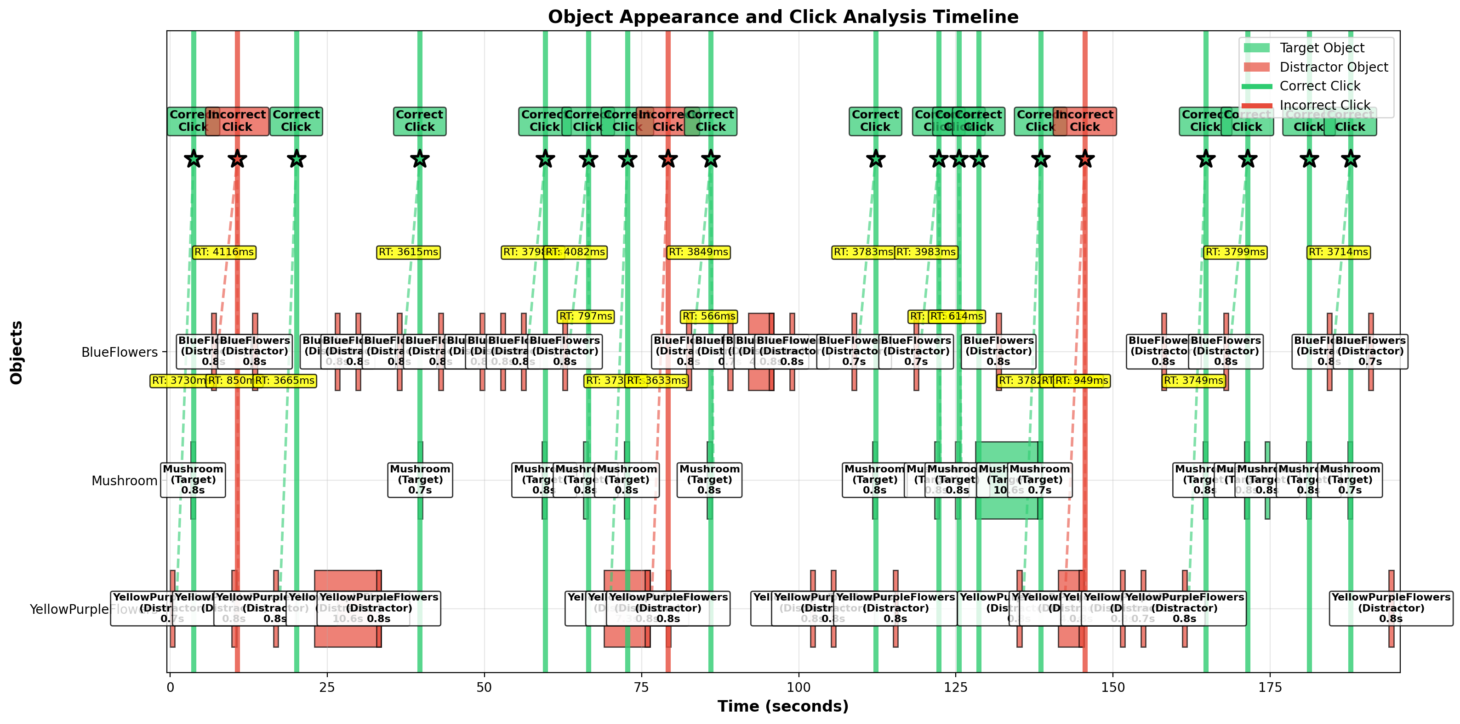}
 \caption{Temporal Analysis of Eye Tracking During Gameplay. The y-axis lists object types (targets and distractors), and the x-axis shows the task timeline. Green bars and lines indicate target appearances and correct clicks, while red/grey bars mark distractors. Red stars denote incorrect clicks on distractors.}
 \label{Temporal Analysis of Eye Tracking During Gameplay}
\end{figure*}

Fig. \ref{Temporal Analysis of Eye Tracking During Gameplay} provides a detailed chronological overview of a user’s performance in an attention game. The Y-axis categorizes the objects into three types: Mushrooms (the target), Blue Flowers, and Yellow-Purple Flowers (both distractors), while the X-axis represents the timeline of the task, spanning approximately 196.7 seconds at the first level of the gameplay. Each object’s appearance is marked by a horizontal bar, with green bars indicating the presence of target objects and red or grey bars representing distractors. Vertical green lines signify when target objects appear, and green stars labeled "Correct Click" indicate moments when the participant successfully identified and clicked on a target. In contrast, a single red star labeled "Incorrect Click" appears around 3 times, highlighting the only instance where the participant mistakenly clicked on a distractor.
During the gameplay, the correct click is annotated with a yellow box displaying the reaction time (RT) in milliseconds, which shows the time it took the user to respond after the target appeared. Reaction times vary considerably, ranging from as fast as 566 milliseconds to around 3900 milliseconds, suggesting variability in attention or response readiness. Over time, RTs improve, indicating potential adaptation or learning as the task progresses. Despite frequent and evenly spaced appearances of distractors, the participant avoids nearly all of them, making just one incorrect click, which demonstrates strong attentional control and task focus. In total, 16 target objects appear during the task, and the user correctly clicks on most of them, reflecting a very high level of accuracy. Each target object remains visible for about 0.8 seconds, during which the participant must respond quickly to register a hit. Distractors appear more often and last for similar durations, adding to the challenge of the task. The single mistake and a few slower reactions show some occasional lapses, but these do not significantly affect the overall firm performance.

\begin{figure}[!ht]
 \centering
 \includegraphics[trim={0 0 0 1.9cm},clip,width=\linewidth]{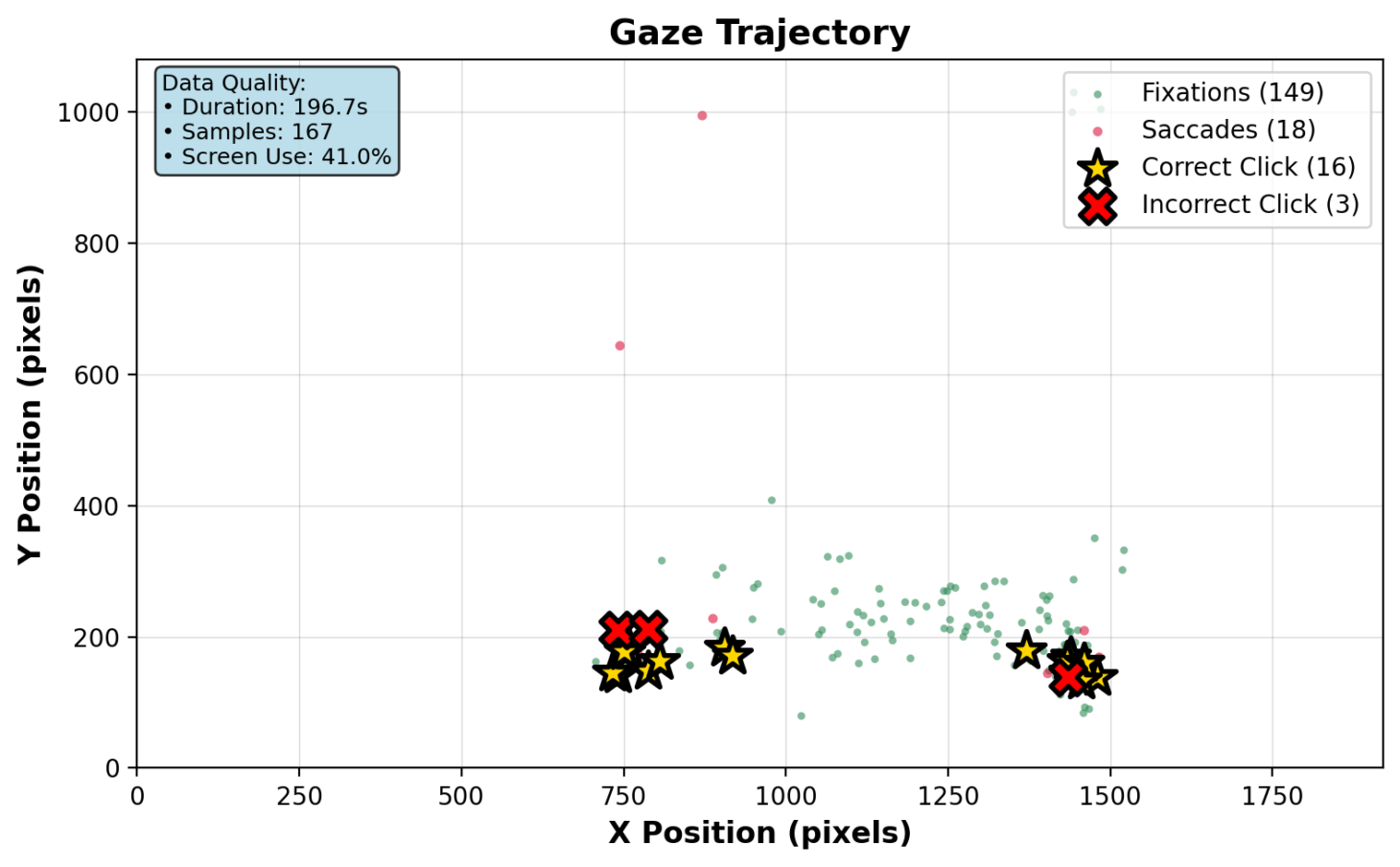}
 \caption{Eye Movement pattern Analysis. The pink dots indicate the saccades (rapid eye movements between fixations), and the light green dots represent the fixations made by the students. The yellow stars indicate correct clicks, and the red cross indicates incorrect clicks. }
 \label{Eye Movement pattern Analysis}
\end{figure}

Fig. \ref{Eye Movement pattern Analysis} presents the eye movement patterns, where it can be observed that students mostly looked into the area of interest (AoI). A total of 149 fixations were recorded, suggesting relatively active visual scanning behavior. Additionally, 33 Saccades were recorded, which typically depict short movements; their presence between fixation points indicates natural eye movement dynamics. Large yellow stars indicate 16 correct clicks, suggesting that the user accurately identified and selected target elements on the screen. These correct clicks are organized into two main clusters, one on the left side and another on the right side of the screen, indicating where the targets were likely located. A single red X represents the only incorrect click, positioned near a group of correct clicks, which may indicate a near-miss error. The total duration of the task was 196.7 seconds, during which 167 gaze samples were recorded. The user engaged with only 4\% of the screen area, suggesting that their focus was on specific regions rather than the entire interface. Overall, the user demonstrated a high level of accuracy, successfully selecting 15 out of 16 targets, corresponding to an accuracy rate of approximately 93.8\%.

\begin{figure}[!ht]
 \centering
 \includegraphics[trim={0 0 0 1.9cm},clip,width=\linewidth]{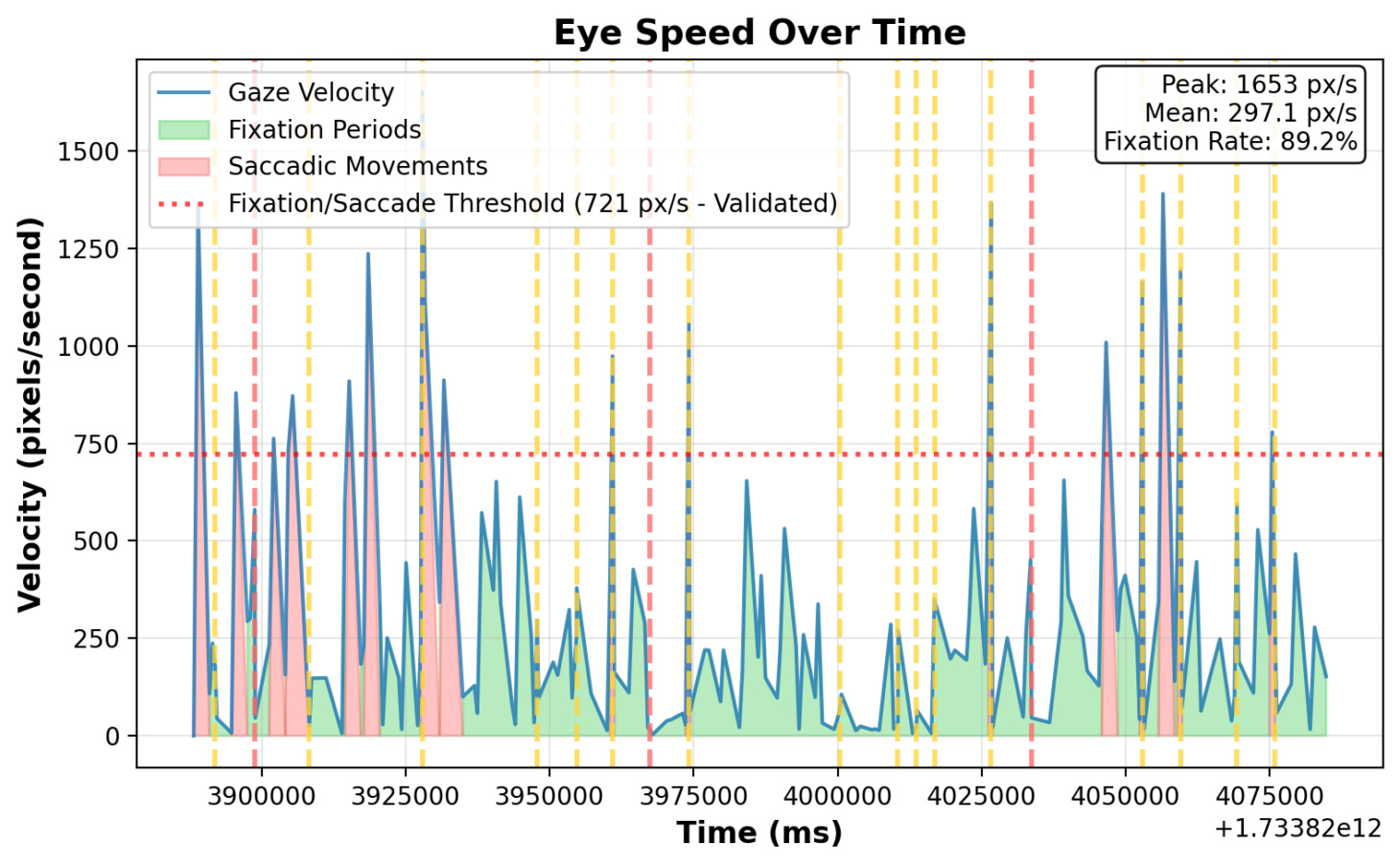}
 \caption{Eye Speed Over Time. The x-axis shows time (ms) and the y-axis shows gaze velocity (px/s). The blue line represents gaze velocity over time. Green areas indicate fixations (slow eye movement), red areas show saccades (rapid shifts), and the red dashed line marks the threshold distinguishing the two.}
 \label{Eye Speed Overtime}
\end{figure}

Fig. \ref{Eye Speed Overtime} illustrates the gaze velocity of students during Level 1 gameplay. The x-axis represents time in milliseconds, while the y-axis shows velocity in pixels per second, reflecting how quickly a user’s gaze moves across the screen. The blue line tracks gaze velocity, capturing fluctuations in eye movement speed from moment to moment. Green-shaded areas indicate fixation periods, during which eye movement slows below a validated threshold, suggesting that the user is likely focusing on a specific area. In contrast, red-shaded regions highlight saccadic movements, where the eye rapidly shifts between different points of interest. A red dashed line marks the fixation/saccade threshold at 721 pixels per second, distinguishing between slower fixation movements and faster saccades. Peaks in the velocity curve that exceed this line are interpreted as saccades. During the session, the peak velocity reached 1643 pixels per second, while the average velocity was 297.1 pixels per second. The fixation rate was 89.2\%, indicating that most of the session was spent in visual fixation rather than scanning. Additionally, vertical dashed lines represent different click events, such as 'Correct', 'Incorrect', or 'Neutral'.

\begin{figure*}[!ht]
 \centering
 \includegraphics[trim={0 0 0 2cm},clip,width=0.8\linewidth]{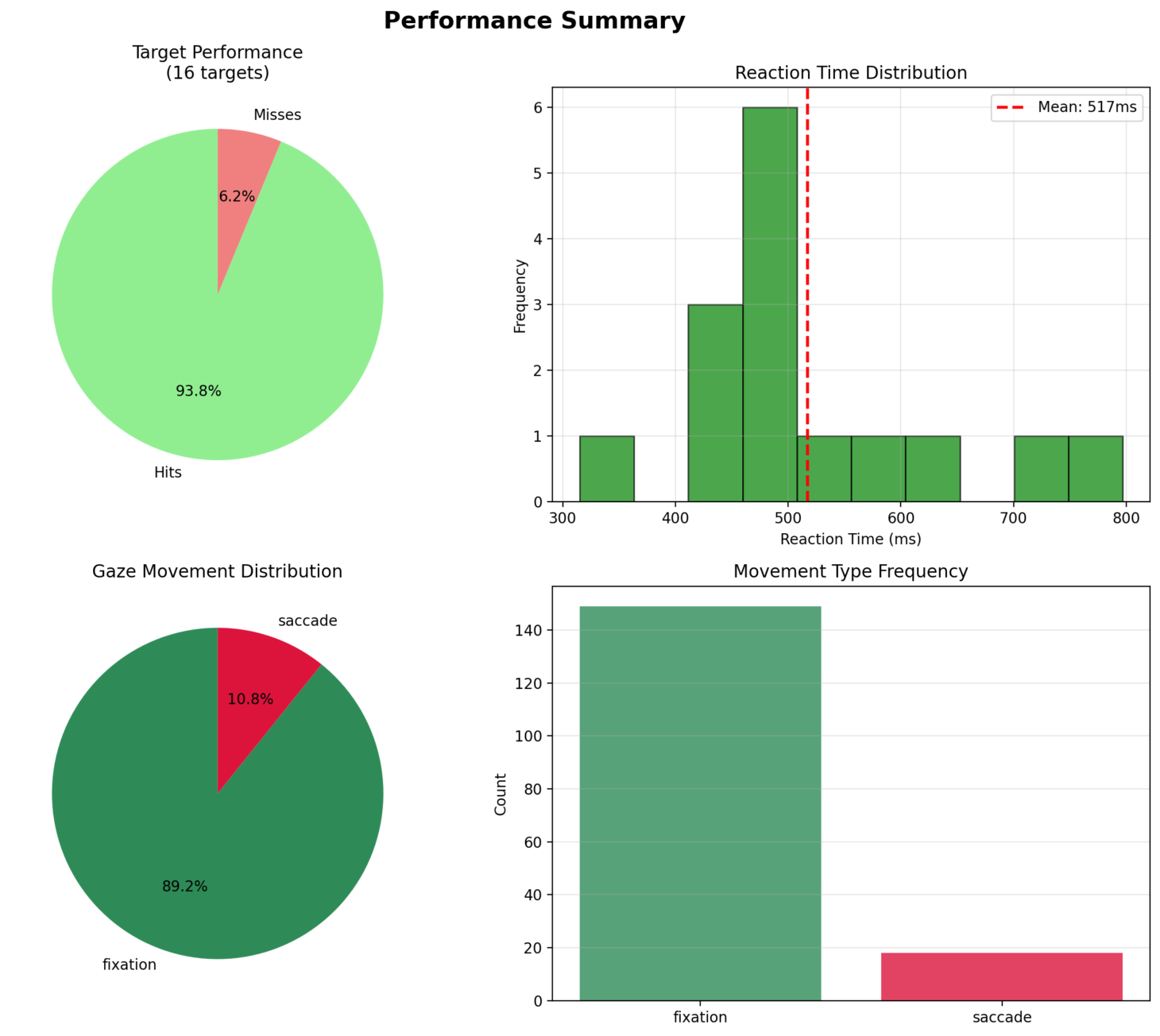}
 \caption{Performance Summary }
 \label{Performance Summary}
\end{figure*}

Fig. \ref{Performance Summary} provides a comprehensive overview of the students' target detection accuracy, reaction speed, gaze movement, and movement frequency at level 1 of the gameplay. In the top-left pie chart, the student demonstrated perfect accuracy by hitting 15 targets, resulting in 93.8\% hits and 6.2\% misses. The second graph (top-right bar graph) shows that the majority of the students' reaction times were between 400 and 550 milliseconds, with a calculated average of 517 ms. This suggests consistent and relatively quick responses. The 3rd graph, bottom-left pie chart, reveals that 89.2\% of the user’s eye movements were fixations, steady visual focus, while only 10.8\% were saccades, which are rapid eye movements between points. Lastly, the fourth graph, bottom-right bar graph, visually reinforces this finding, showing that fixations occurred more than 140 times, whereas saccades were recorded only about 20 times. Together, these graphs indicate not only high task accuracy and efficient response times but also a gaze pattern dominated by stable visual attention.

\begin{figure*}[!ht]
     \centering
     \includegraphics[trim={0 0 0 1cm},clip,width=\linewidth]{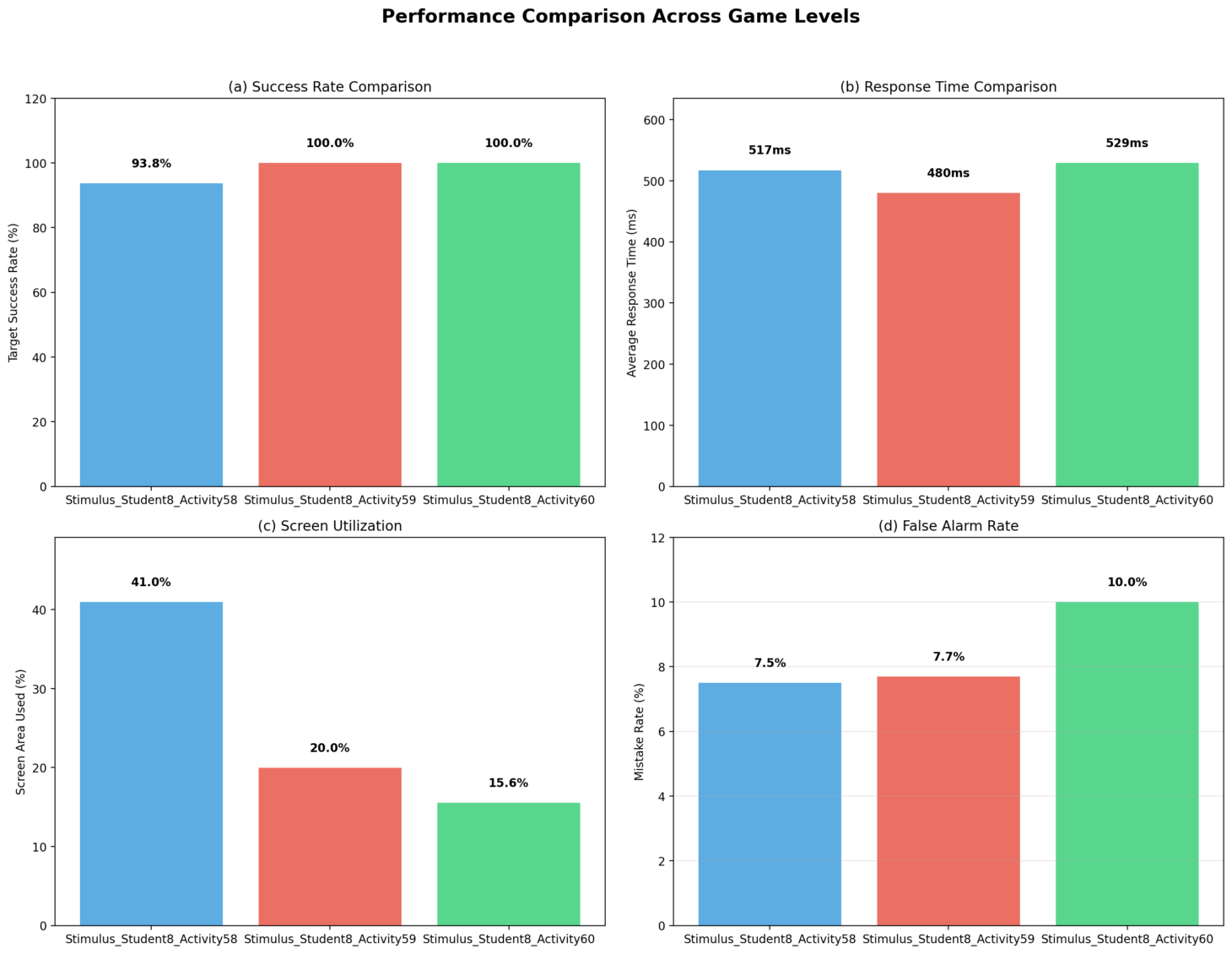}
 \caption{Multilevel Performance Comparison}
 \label{Multilevel Performance Comparison}
\end{figure*}

Fig. \ref{Multilevel Performance Comparison} depicts 4 graphs. The first graph compares the success rates, the second graph shows the response times, the third graph displays the total screen area used, and the fourth graph illustrates the number of mistakes made. It can be seen that the highest target hit rate was achieved at the 2nd level, and the lowest was at the 1st level. The highest time response time was on level 1, and the lowest was on level 2. Furthermore, the student occupies the most screen area at level 1 and is the weakest on level 3. Finally, the student made most of the mistakes in level 3.
 

The results reveal a clear picture of this student's attention and learning patterns. In level 1, they struggled a little, finding around 94\% of targets with slow responses (517 ms), indicating that they were still learning the task. Levels 2 and 3 are the best performers, with a 100\% success rate and significantly faster responses (480 ms at Level 2), demonstrating their ability to perform well when focused. However, level 3 showed a concerning increase in response time to 529 ms, suggesting that the participant may have become tired, overconfident, or perhaps the task had become too complex. The student also spent less time on the screen over time (from 41\% to 15.6\%), which indicates that the student was highly focused on the AoI area. It can also be observed that the number of mistakes made by students increases as the level of difficulty increases.
 
For psychologists, this data provides objective evidence of attention patterns that they can precisely observe when attention peaks and declines, measure how long the student can maintain focus, and track whether interventions or medications are effective by comparing these numbers over time. Teachers can use this information to time their most important lessons during the student's peak attention period (such as level 2), plan breaks before fatigue sets in (before level 3 declines), and recognize that this student requires extra support with impulse control, as their mistake rate increases dramatically when they are tired. Both professionals can work together using these concrete numbers to create personalized strategies that align with the student's natural attention rhythms.

\section{Conclusion and Future Work}\label{conclusion}
This paper presented a time-series analysis tool that combines multiple analytical dimensions: enhanced movement classification that distinguishes between fixations, saccades, and smooth pursuit; temporal pattern analysis that reveals how attention evolves throughout task performance; object-click sequence tracking that directly links visual attention to user actions; and performance metrics that quantify both accuracy and efficiency of visual search behavior. Psychologists and teachers can collaborate using this tool to create personalized strategies that align with the student's natural attention rhythms. In the future, we plan to design Large Language Models (LLMs) to enhance this eye-tracking tool by making it more intelligent and user-friendly (e.g., LLMs can help automatically label gaze patterns, generate easy-to-understand summaries of user behavior, and allow researchers to interact with the tool using simple language commands. This means users could ask questions like "When was the user most distracted?" and get helpful insights without needing complex code or manual analysis.)
 
\section{Acknowledgment}\label{ack}
The research leading to these results is within the frame of the "EMPOWER. Design and evaluation of technological support tools to empower stakeholders in digital education" project, which has received funding from the European Union’s Horizon Europe program under grant agreement No 101060918. Views and opinions expressed are, however, those of the author(s) only and do not necessarily reflect those of the European Union. Neither the European Union nor the granting authority can be held responsible for them.

\bibliographystyle{tfnlm}
\bibliography{ref-time}

\begin{thebibliography}{10}
\providecommand{\url}[1]{\normalfont{#1}}
\providecommand{\urlprefix}{Available from: }

\bibitem{rehman2024towards}
Rehman~A, Heldal~I, Stilwell~D, et~al. Towards a supporting framework for neuro-developmental disorder: Considering artificial intelligence, serious games and eye tracking. In: 2024 IEEE International Conference on Big Data (BigData); IEEE; 2024. p. 8238--8240.

\bibitem{novak2024eye}
Nov{\'a}k~J{\v{S}}, Masner~J, Benda~P, et~al. Eye tracking, usability, and user experience: A systematic review. International Journal of Human--Computer Interaction. 2024;\hspace{0pt}40(17):4484--4500.

\bibitem{ali2023towards}
Ali~Q, Heldal~I, Helgesen~CG. Towards a framework for visualization and analysis of eye tracking data for functional vision screening. In: Proceedings of the 3rd International Health Data Workshop (HEDA 2023); (CEUR Workshop Proceedings; Vol. 3440). CEUR-WS.org; 2023. p. 3--12. \urlprefix\url{https://ceur-ws.org/Vol-3440/paper3.pdf}.

\bibitem{daehlen2024towards}
D{\ae}hlen~A, Heldal~I, Rehman~A, et~al. Towards more accurate help: Informing teachers how to support ndd children by serious games and eye tracking technologies. In: Proceedings of the 2024 Symposium on Eye Tracking Research and Applications; 2024. p. 1--7.

\bibitem{costescu2023mushroom}
Costescu~C, David~C, Roșan~A, et~al. Mushroom hunters: A digital game for assessing and training sustained attention in children with neurodevelopmental disorders. In: International Conference in Methodologies and intelligent Systems for Techhnology Enhanced Learning; Springer; 2023. p. 78--86.

\bibitem{lamsa2022focus}
L{\"a}ms{\"a}~J, Kotkajuuri~J, Lehtinen~A, et~al. The focus and timing of gaze matters: Investigating collaborative knowledge construction in a simulation-based environment by combined video and eye tracking. In: Frontiers in education; Vol.~7; Frontiers Media SA; 2022. p. 942224.

\bibitem{keshava2024just}
Keshava~A, Nezami~FN, Neumann~H, et~al. Just-in-time: Gaze guidance in natural behavior. PLOS Computational Biology. 2024;\hspace{0pt}20(10):e1012529.

\bibitem{papavlasopoulou2021investigating}
Papavlasopoulou~S, Sharma~K, Melhart~D, et~al. Investigating gaze interaction to support children’s gameplay. International Journal of Child-Computer Interaction. 2021;\hspace{0pt}30:100349.

\bibitem{frutos2015assessing}
Frutos-Pascual~M, Garcia-Zapirain~B. Assessing visual attention using eye tracking sensors in intelligent cognitive therapies based on serious games. Sensors. 2015;\hspace{0pt}15(5):11092--11117.

\bibitem{piazzalunga2023investigating}
Piazzalunga~C, Dui~LG, Termine~C, et~al. Investigating visual perception impairments through serious games and eye tracking to anticipate handwriting difficulties. Sensors. 2023;\hspace{0pt}23(4):1765.

\bibitem{velichkovsky2019visual}
Velichkovsky~BB, Khromov~N, Korotin~A, et~al. Visual fixations duration as an indicator of skill level in esports. In: Human-Computer Interaction--INTERACT 2019: 17th IFIP TC 13 International Conference, Paphos, Cyprus, September 2--6, 2019, Proceedings, Part I 17; Springer; 2019. p. 397--405.

\bibitem{kim2024development}
Kim~M, Lee~J, Lee~SY, et~al. Development of an eye-tracking system based on a deep learning model to assess executive function in patients with mental illnesses. Scientific Reports. 2024;\hspace{0pt}14(1):18186.

\bibitem{argasinski2017patterns}
Argasi{\'n}ski~JK, Grabska-Gradzi{\'n}ska~I. Patterns in serious game design and evaluation application of eye-tracker and biosensors. In: International Conference on Artificial Intelligence and Soft Computing; Springer; 2017. p. 367--377.

\bibitem{hajari2018spatio}
Hajari~N, He~W, Cheng~I, et~al. Spatio-temporal eye gaze data analysis to better understand team cognition. In: Smart Multimedia: First International Conference, ICSM 2018, Toulon, France, August 24--26, 2018, Revised Selected Papers 1; Springer; 2018. p. 39--48.

\bibitem{du2024privategaze}
Du~L, Jia~J, Zhang~X, et~al. Privategaze: Preserving user privacy in black-box mobile gaze tracking services. Proceedings of the ACM on Interactive, Mobile, Wearable and Ubiquitous Technologies. 2024;\hspace{0pt}8(3):1--28.

\bibitem{paskovske2024eye}
Paskovske~A, Kliziene~I. Eye tracking technology on children's mathematical education: systematic review. In: Frontiers in Education; Vol.~9; Frontiers Media SA; 2024. p. 1386487.

\bibitem{bueno2023datasets}
Bueno~ML, Thill~S. Datasets for artificial intelligence in education: The case of children with neurodevelopmental disorders. In: International Conference in Methodologies and intelligent Systems for Techhnology Enhanced Learning; Springer; 2023. p. 70--77.

\bibitem{thill2022modelling}
Thill~S, Charisi~V, Belpaeme~T, et~al. From modelling to understanding children's behaviour in the context of robotics and social artificial intelligence. arXiv preprint arXiv:221011161. 2022;\hspace{0pt}.

\bibitem{costescu2020development}
Costescu~C, Rosan~A, Petru~A, et~al. Development of a technological screening platform for children. In: 2020 11th IEEE International Conference on Cognitive Infocommunications (CogInfoCom); IEEE; 2020. p. 000453--000458.

\bibitem{salvucci2000ivt}
Salvucci~DD, Goldberg~JH. Identifying fixations and saccades in eye-tracking protocols. In: Proceedings of the Symposium on Eye Tracking Research \& Applications (ETRA); 2000. p. 71--78.

\bibitem{olsen2012tobii}
Olsen~A. The tobii i-vt fixation filter. Tobii Technology. 2012;\hspace{0pt}21(4-19):5.

\bibitem{CCPT}
PhD~JFS, MSc~MST, MSc~DISW, et~al. Sustained attention and executive functioning performance in attention-deficit/hyperactivity disorder. Child Neuropsychology. 2005;\hspace{0pt}11(3):285--294. PMID: 16036452; \urlprefix\url{https://doi.org/10.1080/09297040490916938}.

\bibitem{holmqvist2011eye}
Holmqvist~K, Nystr{\"o}m~M, Andersson~R, et~al. Eye tracking: A comprehensive guide to methods and measures. oup Oxford; 2011.

\bibitem{henderson2003human}
Henderson~JM. Human gaze control during real-world scene perception. Trends in Cognitive Sciences. 2003;\hspace{0pt}7(11):498--504.

\bibitem{nyström2010detection}
Nystr{\"o}m~M, Andersson~R, Holmqvist~K, et~al. The influence of calibration method and eye physiology on eyetracking data quality. Behavior Research Methods. 2010;\hspace{0pt}42(1):188--204.

\bibitem{jarodzka2010eye}
Jarodzka~H, Holmqvist~K, Gruber~H. Eye tracking and education: A selective review of the literature. Educational Psychology Review. 2010;\hspace{0pt}22(2):123--146.

\end{thebibliography}
\end{document}